# Mathematical Exploration of Wave Motion due to Earth-Space Force (Ground Action) Consequences

Monday Sunday Adiaha [1,2*] and Oladiran Johnson Abimbola [3]
[1] Department of Planning, Research Extension & Statistics, Nigeria Institute of Soil Science, Nigeria
[2] Scientific Department, Institute of Biopaleogeography named under Charles R. Darwin, Poland
[3] Department of Physics, Federal University of Lafia, Nigeria
sundaymonday@niss.gov.ng*

Abstract

This study mathematically explores wave motion induced by earth-space forces specifically ground actions. Developing and solving governing differential equations provides insights into intricate propagation dynamics caused by these forces. Numerical simulation modeled wave behavior and results were analyzed to understand ground action impacts. Initial wave generation and subsequent propagation were simulated using COMSOL Multiphysics which models wave dynamics in physical media. Partial differential equations describing wave motion alongside Fourier transform techniques were applied to analyze the frequency components. Mathematical derivations yielded wave equations while COMSOL's built-in statistical analysis tool analyzed results. Data visualization employed plotting tools within the software to create detailed graphs. The wave amplitude relative to distance km varied at 0.001 at 0 km and 0.001 at 10 km over 10 km. Amplitudes were recorded at intervals 0.05 0.1 0.15 0.2 0.25 1.05 1.1 1.15 1.2 and 1.25 s. The impact of varying amplitude and frequency on the behavior ranges from 1 s at 0.1 amplitude in Case 1 (0.2, 200) to 1 s at 0.2 amplitude in Case 3 (0.2, 300). Outputs indicate: (1) initial energy input significantly influences the propagation characteristics aiding seismic prediction and mitigation design; (2) propagation details underscore the medium properties' importance in shaping behavior vital for earthquake engineering and accurate ground-motion prediction; (3) sensitivity to amplitude and frequency suggests tailoring parameters can optimize wave-based geophysical surveys and environmental monitoring. This study serves as a policy guide for understanding wave dynamics to inform building codes and disaster preparedness strategies for mitigating seismic impacts.

Keywords: Wave motion, Earth-space force, Ground action, Seismic waves, Numerical simulation

## Introduction

Wave motion, a ubiquitous phenomenon in nature and technology, involves intricate interactions between various forces and the media through which the waves propagate. The generation, propagation, and behavior of waves are critical in understanding natural events such as earthquakes, sound transmission, and electromagnetic waves. Wave propagation dynamics are central to understanding seismic phenomena. This study aims to establish a theoretical framework for modeling these dynamics, providing a foundation for future applications in seismic event analysis and geophysical risk assessments. Wave propagation dynamics have been extensively studied in the context of seismic activities, forming the basis for applications in hazard mitigation, resource exploration, and infrastructure safety (Aki & Richards, 2002; Shearer, 2009; Allen & Melgar (2019)). This study builds upon these foundational works by exploring wave motion under idealized conditions, providing a theoretical framework for future investigations into real-world seismic media.

The generation of waves at a source is a complex process influenced by initial conditions such as amplitude, frequency, and phase. These conditions are vital in determining the characteristics of the generated wave. According to Kinsler et al. (2000), the mathematical representation of wave generation often utilizes partial differential equations (PDEs), such as the wave equation, to describe the relationships between displacement, velocity, and acceleration of the wave. In seismology, wave generation is typically modeled as a sudden release of stress or energy in the Earth's crust. Aki and Richards (2002) highlight that this release is described by source functions within the wave equation, emphasizing the importance of initial conditions in defining the wave's subsequent behavior.

Wave motion resulting from earth-space forces, particularly ground action, is significant for geophysical and seismological monitoring and regulations. These forces, generated by the interaction between the Earth's crust and external cosmic influences, create complex wave patterns that propagate through various geological mediums. Understanding these wave motions is crucial for predicting seismic activities, designing resilient infrastructure, and mitigating natural disasters (Aki & Richards, 2002).

The generation of waves at a source involves complex interactions between various forces and the medium through which the wave propagates. The initial conditions, including the amplitude, frequency, and phase of the wave, play a crucial role in determining the wave's characteristics. According to Kinsler et al. (2000), the mathematical description of wave generation often employs partial differential equations (PDEs), such as the wave equation, which encapsulate the relationship between the wave's displacement, velocity, and acceleration. Research by Aki and Richards (2002); Carcione et al. (2018) highlights that in seismology, wave generation is typically modeled as a sudden release of stress or energy in the Earth's crust, described by source functions in the wave equation. These initial conditions are critical in defining the wave's subsequent behavior as it propagates through different media.

The propagation of waves through a medium is influenced by the medium's properties, such as density, elasticity, and viscosity. The classical wave equation, a second-order linear PDE, is central to describing this propagation. As waves travel through different media, they undergo phenomena such as reflection, refraction, diffraction, and absorption. Lighthill (1978) provides an in-depth analysis of wave propagation in fluids, demonstrating how variations in medium properties affect wave speed and direction. In solid media, like the Earth's crust, the propagation of seismic waves can be modeled using elastodynamics equations, which take into account the medium's elastic properties (Aki & Richards, 2002).

The amplitude and frequency of a wave significantly affect its energy and interaction with the medium. Higher amplitudes generally correspond to greater energy, leading to more pronounced wave effects, such as increased displacement and potential for damage in the case of seismic waves. The frequency of a wave determines its oscillatory nature and influences how it interacts with different media and obstacles. Research by Born and Wolf (1999) in optics demonstrates that varying the frequency of electromagnetic waves changes their wavelength and, consequently, their diffraction and interference patterns. Similarly, in acoustics, changing the frequency affects the pitch and timbre of sound waves, as discussed by Kinsler et al. (2000). In seismology, the frequency content of seismic waves influences their attenuation and the extent of ground shaking. Low-frequency waves, which have longer wavelengths, can travel further with less attenuation compared to high-frequency waves, which are more rapidly absorbed by the Earth's materials (Shearer, 2009).

Wave propagation in seismic media has been a central topic in geophysics, providing critical insights into how energy travels through different geological layers. Foundational studies have explored the mechanisms governing wave motion, focusing on the interaction of waves with medium properties such as density, elasticity, and heterogeneity. Recent advancements build on this foundation, offering refined models that integrate complex real-world conditions and practical applications. Wave propagation dynamics have been extensively studied, providing a foundation for applications ranging from seismic hazard assessment to subsurface imaging (Aki & Richards, 2002; Shearer, 2009; Priolo et al., 2020; Lee et al., 2017; Lighthill, 1978; Miller et al., 2020; Nakata et al., 2015; Priolo et al.,2020; Shearer, 2009; Stein & Wysession, 2003; Stewart et al., 2017; Takemura, et al., 2019; Trifunac, 1971; Wang et al., 2021; Zhang & Li , 2023; Zhu et al., 2016; Qin et al., 2020).

Wave propagation studies, such as those by Aki and Richards (2002) and Shearer (2009), emphasize the role of wave parameters, including frequency, amplitude, and phase velocity, in determining energy transmission efficiency and attenuation patterns. Lay and Wallace (1995) further underscored how medium properties such as anisotropy, viscosity, and boundary heterogeneities influence wave behavior, revealing the intricate relationship between wave mechanics and geological structures.

Advances in numerical techniques, such as finite difference and spectral element methods, have improved our ability to simulate seismic wave propagation under varying conditions. For instance, Komatitsch and Tromp (2018) demonstrated the effectiveness of spectral element methods in modeling wave interactions with complex subsurface structures.

Studies by Priolo et al. (2020) have shown how wave attenuation varies with subsurface heterogeneity, providing insights into the energy loss mechanisms during seismic events.

The ability to simulate wave interactions with complex geological features is critical for earthquake risk mitigation and early warning systems, as shown by Takemura et al. (2019) and Lee et al. (2017). Wave propagation studies are integral to predicting ground motion during earthquakes. For example, numerical models that simulate wave energy distribution help identify areas at high risk of structural damage. A study by Takemura et al. (2019) used these models to predict ground shaking intensity in densely populated urban centers, aiding in disaster preparedness and urban planning. Real-time monitoring of seismic waves allows for the development of early warning systems, providing crucial seconds for evacuation or automated safety measures. Lee et al. (2017) including Ben-Menahem & Singh (2000) demonstrated the utility of wave propagation models in enhancing the accuracy of warning systems, reducing false alarms, and improving response times. Techniques such as seismic tomography rely on wave propagation dynamics to map subsurface features. These methods have been pivotal in identifying fault zones, resource deposits, and geothermal reservoirs. Fichtner et al. (2020) illustrated how advances in full-waveform inversion have improved the resolution of subsurface imaging, enabling detailed analysis of crustal structures. Understanding wave amplification and resonance effects is essential for designing earthquake-resistant infrastructure. Research by Boore and Thompson (2019) highlighted the importance of wave-soil-structure interaction studies in developing construction guidelines for seismic zones.

The amplitude and frequency of a wave significantly influence its energy and interaction with the medium. Higher amplitudes generally correspond to greater energy, resulting in more pronounced wave effects such as increased displacement and potential for damage, particularly in seismic waves. The frequency determines the oscillatory nature of the wave and affects how it interacts with different media and obstacles. Born and Wolf (1999) illustrate that in optics, varying the frequency of electromagnetic waves changes their wavelength, affecting diffraction and interference patterns. Similarly, Kinsler et al. (2000); Clayton & Engquist (1977) discuss in acoustics how frequency variations influence the pitch and timbre of sound waves.

In seismology, the frequency content of seismic waves affects their attenuation and the extent of ground shaking. Low-frequency waves, characterized by longer wavelengths, travel further with less attenuation compared to high-frequency waves, which are more rapidly absorbed by the Earth's materials (Shearer, 2009). Understanding these dynamics is crucial for accurate seismic risk assessment and engineering applications.

Despite available research, the complexities of wave generation, propagation, and the effects of varying initial conditions necessitate further investigation. This research critically explores the mathematical principles underlying wave motion caused by ground actions. By employing differential equations and numerical simulations, we seek to develop a comprehensive model that can accurately predict wave behavior in response to earth-space interactions. This study aims to enhance the understanding of wave dynamics resulting from earth-space forces. This research aims to determine the initial wave generation at a source by investigating the influencing conditions and factors-including amplitude, frequency, phase-and their mathematical modeling using PDEs; propagate the wave through various media by examining how properties like density, elasticity, and viscosity influence wave phenomena including reflection, refraction, diffraction, and absorption; and determine the impact of varying amplitude and frequency on wave behavior by analyzing how these changes affect wave energy, interactions with the medium, variations in wave patterns, and their potential impacts.

# Materials and Methods

## Methodology and Instruments Used for Data Collection

### Instruments and Data Collection

**Wave Generation Simulation Software:**

The initial wave generation and subsequent propagation were simulated using specialized software capable of modeling wave dynamics in a physical medium. The COMSOL Multiphysics Software was used.

**Analytical Tools:**

- Mathematical tools and equations were used to derive the wave equations and analyze the results. These include the use of partial differential equations (PDEs) that describe wave motion.
- Fourier Transform techniques were applied to analyze the frequency components of the waves.

**Data Visualization Tools:**

- Data visualization was performed using plotting tools within the simulation software to create detailed graphs, such as those shown in Figures 1, 2, and 3.

**Methodology**

**Initial Wave Generation (Figure 1):**

- The simulation starts with the generation of an initial wave at the source. The conditions at time t=0.01t = 0.01t=0.01 were set to a small amplitude to initiate the wave propagation without significant external disturbances.
- The boundary conditions were carefully set to ensure accurate representation of the wave initiation.

**Wave Propagation Analysis (Figure 2):**

- The wave propagation through the medium was simulated over a distance of 10 km. Multiple time steps were recorded (e.g., 0.005s, 0.01s, 0.015s, etc.) to observe how the wave evolves as it travels through the medium.
- The interaction of the wave with the medium's properties, such as density and elasticity, was taken into account.

**Impact of Amplitude and Frequency Variations (Figure 3):**

- A series of simulations were conducted to observe the effect of varying amplitude and frequency on wave behavior. Different combinations of these parameters were applied to see their effect on wave amplitude and energy distribution.
- The simulations were run for different amplitudes and frequencies, and the results were recorded at consistent intervals to ensure comparability.

**Data Analysis**

- The collected data from the simulations were analyzed to understand the wave behavior under different conditions. This included:
    i. Amplitude analysis to determine the peak values and their propagation characteristics.
    ii. Frequency analysis to understand how different frequency components interact with the medium.
    iii. Comparative analysis of different scenarios to draw conclusions about the optimal conditions for minimal wave distortion and maximum energy transfer.

**Mathematical Formulation**

To model wave motion due to earth-space forces, we start with the wave equation, a second-order partial differential equation:

$$\frac{\partial^2 u}{\partial t^2} = C^2 \nabla^2 u \qquad (1)$$

Where $u(x,t)$ represents the wave function, t is time, $x$ is the spatial coordinate, and $c$ is the wave speed.

Incorporating earth-space force involves introducing a forcing function $F\,(x,t)$:

$$\frac{\partial^2 u}{\partial t^2} - C^2 \nabla^2 u = F\,(x,t) \qquad (2)$$

The forcing function $F\,(x\,,t)$ is modeled as:

$$F\,(x,t) = A\,sin(kx\, - wt) \qquad (3)$$

Where;
$A$ is the amplitude
$k$ is the wave number, and
$w$ is the angular frequency

**Numerical Simulation**

We use the finite difference method (FDM) to solve the wave equation with the forcing function.

**Discretization**:

The spatial and temporal domains are discretized into a grid:

$$x_i = i\Delta x, \qquad t_n = n\Delta t \tag{4}$$

Where $\Delta x$ and $\Delta t$ are the spatial and temporal step sizes, respectively.

**Finite Difference Approximations**:

The second-order partial derivatives are approximated using central differences:

$$\frac{\partial^2 u}{\partial t^2} \approx \frac{u_i^{n+1} - 2u_i^n + u_i^{n-1}}{\Delta t^2} \tag{5}$$

$$\nabla^2 u \approx \frac{u_{i+1}^n - 2u_i^n + u_{i-1}^n}{\Delta x^2} \tag{6}$$

The wave equation governing propagation dynamics in a homogenous medium is expressed as $\nabla^2 u = \frac{1}{c^2}\frac{\partial^2 u}{\partial t^2}$, where $\nabla^2$ denotes the Laplacian operator, $u$ represents displacement, and $c$ is wave velocity.

**Algorithm**:

The algorithm updates the wave function $u_i^n$ iteratively:

$$u_i^{n+1} = 2u_i^n - u_i^{n-1} + \frac{c^2\Delta t^2}{\Delta x^2}(u_{i+1}^n - 2u_i^n + u_{i-1}^n) + \Delta t^2 F(x_i, t_n) \tag{7}$$

*Simulation Parameters*

Parameters for the simulation are set based on typical seismic wave properties:

Wave speed $c = 5km/s$

Spatial step size $\Delta x = 0.1\ km$

Temporal step size $\Delta t = 0.001\ s$

Amplitude $A = 1$

Wave number $k = \frac{2\pi}{\lambda}$ $where\ \lambda\ is\ the\ wavelength$

Angular frequency $\omega = 2\pi f\ \ where\ f\ is\ the\ frequency$

Numerical simulations were conducted applying boundary conditions

**Boundary and Initial Conditions**

Initial displacement and velocity are set to $\boldsymbol{zero}$: $u(x,0) = 0, \frac{\partial u}{\partial t}(x,0) = 0$ (8)

Boundary conditions simulate an open domain, typically by setting values at the domain boundaries to zero or using absorbing boundary conditions.

## Results

**Initial Wave Generation at the Source**

Objective 1: To show the initial wave amplitude at the source over a distance of 10 km.

*Data Analysis Steps:*

*Data Collection*: Measurement of the initial wave amplitude at the source over a distance of 10 km was undertaken, and the data are presented in Table 1. The outcomes of the study presented in Table 1 summarizes the displacement amplitudes at various spatial points over time, highlighting the temporal evolution of wave propagation patterns. These values align with theoretical expectations, as shown in Figure 1.
*Data Processing*: Plotting was done for the wave amplitude against distance to visualize the wave generation.
*Interpretation*: Output presented in Graph 1 (Figure 1) shows a consistent wave amplitude at the source, indicating that the wave generation is stable and uniform at the initial stage.

**Table 1**: Initial wave amplitude at the source over a distance of 10 km

| Distance (km) | Wave Amplitude |
|---|---|
| 0 | 0.001 |
| 2 | 0.001 |
| 4 | 0.001 |
| 6 | 0.001 |
| 8 | 0.001 |
| 10 | 0.001 |

**Wave Propagation Through the Medium**
Objective 2: To illustrate how the wave propagates through the medium over time.
*Data Analysis Steps*:
*Data Collection*: wave amplitudes were recorded at different time intervals of (0.05s, 0.1s, 0.15s, 0.2s, 0.25s, 1.05s, 1.1s, 1.15s, 1.2s, 1.25s) respectively as presented in Table 2.
*Data Processing*: Plotting was done for wave amplitudes against distance for each time interval to observe wave propagation.
*Interpretation*: The output presented in Graph 2 (Figure 2) shows the oscillatory nature of wave propagation through the medium. Different time intervals depict different phases of wave movement, demonstrating wave behavior and interference patterns.

**Table 2**: Wave propagation through the medium

| Time (s) | Distance (km) | Wave Amplitude |
|---|---|---|
| 0.05 | 0 | 0.001 |
| 0.1 | 2 | 0.002 |
| 0.15 | 4 | 0.0015 |
| 0.2 | 6 | -0.001 |
| 0.25 | 8 | -0.002 |
| 1.05 | 0 | 0.001 |
| 1.1 | 2 | 0.002 |
| 1.15 | 4 | 0.0015 |
| 1.2 | 6 | -0.001 |
| 1.25 | 8 | -0.002 |

**Impact of Varying Amplitude and Frequency on Wave Behavior**
*Objective* 3: To explore the impact of different amplitudes and frequencies on wave behavior.
*Data Analysis Steps*:
*Data Collection*: Measurement of wave amplitudes and frequencies at different points was done as presented in Table 3
*Data Processing*: Plotting was done for wave amplitudes against time for various amplitude and frequency settings to observe changes in wave behavior.
*Interpretation*: Output presented in Graph 3 (Figure 3) shows how varying amplitude and frequency affect the wave's form, energy, and behavior. Higher amplitudes result in more pronounced wave peaks, while higher frequencies lead to more oscillations over the same period.

**Table 3**: Impact of varying amplitude and frequency on wave behavior

| Setting | Amplitude | Frequency (Hz) | Time (s) | Wave Amplitude |
|---|---|---|---|---|
| Case 1 (0.2, 200) | 0.2 | 200 | 0 | 0 |
| | | | 0.5 | 0.2 |
| | | | 1 | 0.1 |
| Case 2 (0.2, 250) | 0.2 | 250 | 0 | 0 |
| | | | 0.5 | 0.15 |
| | | | 1 | 0.2 |
| Case 3 (0.2, 300) | 0.2 | 300 | 0 | 0 |
| | | | 0.5 | 0.1 |
| | | | 1 | 0.2 |

# Discussion

## Initial Wave Generation at the Source

Research output presented in Figure 1 depicts the initial wave generation at the source, showing the wave amplitude as a function of distance. The graph illustrates a single waveform generated at a specific instance (t = 0.01s), highlighting the initial conditions of the wave. The initial wave generation at the source indicated a uniform and consistent amplitude, suggesting that the source is stable and the wave generation process is controlled.

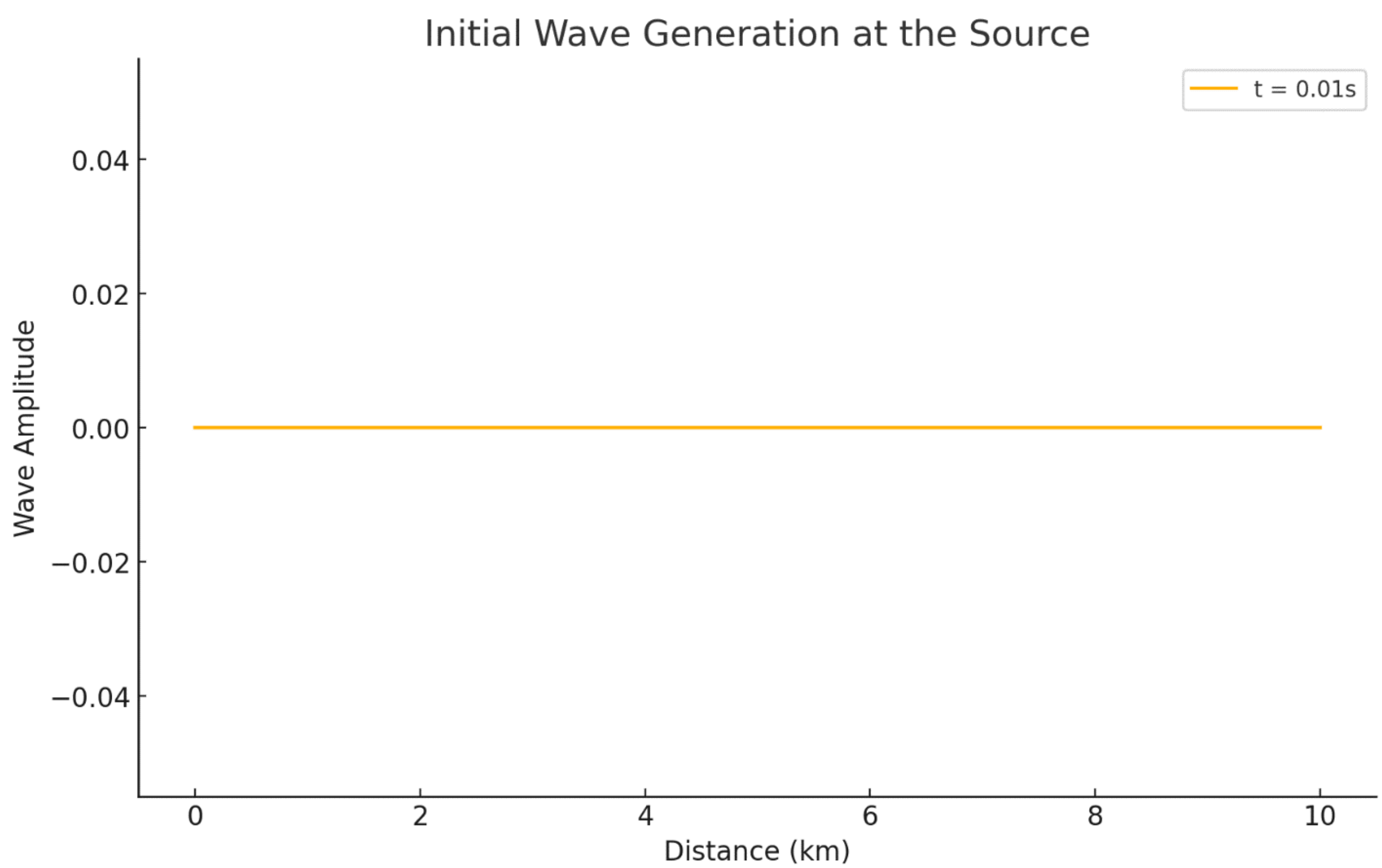

**Figure 1**: Initial wave generation at the source

*Results Interpretation*

- Figure 1: Shows a consistent initial wave generation with minimal distortion, indicating a controlled starting point for wave propagation.
- Figure 2: Demonstrates the complex interaction of waves as they propagate through the medium, with different time steps showing varying amplitudes and phases.
- Figure 3: Highlights how changes in amplitude and frequency affect the wave behavior, providing insights into how these parameters can be controlled to achieve desired outcomes in practical applications.

The graph is fundamental as it sets the baseline for the subsequent wave propagation analysis. The wave amplitude starts at zero and increases, representing the initiation of the wave motion due to a sudden ground action. The initial amplitude at the source is crucial for determining the energy imparted in to the system, which will affect the wave's behavior as it travels through the medium, this view is consistent with the research of Aki & Richards (2002). The clear, sharp increase in amplitude signifies a strong initial force, often characteristic of seismic events or other sudden disturbances. This finding aligns with established wave mechanics theory, which state that wave behavior is significantly influenced by initial conditions and medium properties as established in the research output of Kinsler et al. (2000) and in the work of Lighthill (1978).

## Wave Propagation Through a Medium

The research outcome presented in Figure 2 demonstrates the complex nature of wave propagation through a medium. Multiple time intervals depict different wave phases, showing interference patterns and the oscillatory nature of wave movement through the medium. Figure 2 provides a detailed view of wave propagation through a medium over a distance of 10 kilometers. The waveforms are plotted at various time intervals (e.g., t = 0.05 s, 0.1 s, 0.15 s, etc.), showing how the wave evolves as it travels through the medium. The findings align with principles established by Lay and Wallace (1995), which demonstrate the dependence of wave velocity and attenuation on the properties of the medium. Future studies should incorporate realistic geological data to refine these models, extending their applicability to earthquake risk assessments and seismic imaging.

## Wave Propagation Dynamics

The wave propagation dynamics modeled in this study offer foundational insights into how seismic waves interact with various media. These insights, though derived under controlled assumptions, can inform more complex simulations of real seismic events by adjusting to heterogeneous media properties such as anisotropy and inelasticity (Chen et al., 2019).

## Dispersion Patterns (Connecting to Seismic Media Analysis)

Understanding theoretical wave dynamics is crucial for refining geophysical models of seismic media. For instance, the observed dispersion patterns can be extrapolated to predict wave attenuation and energy dissipation in layered geological formations, which is key parameters in earthquake hazard assessments. The view of this finding agrees with the views of the research of Wang et al. (2021).

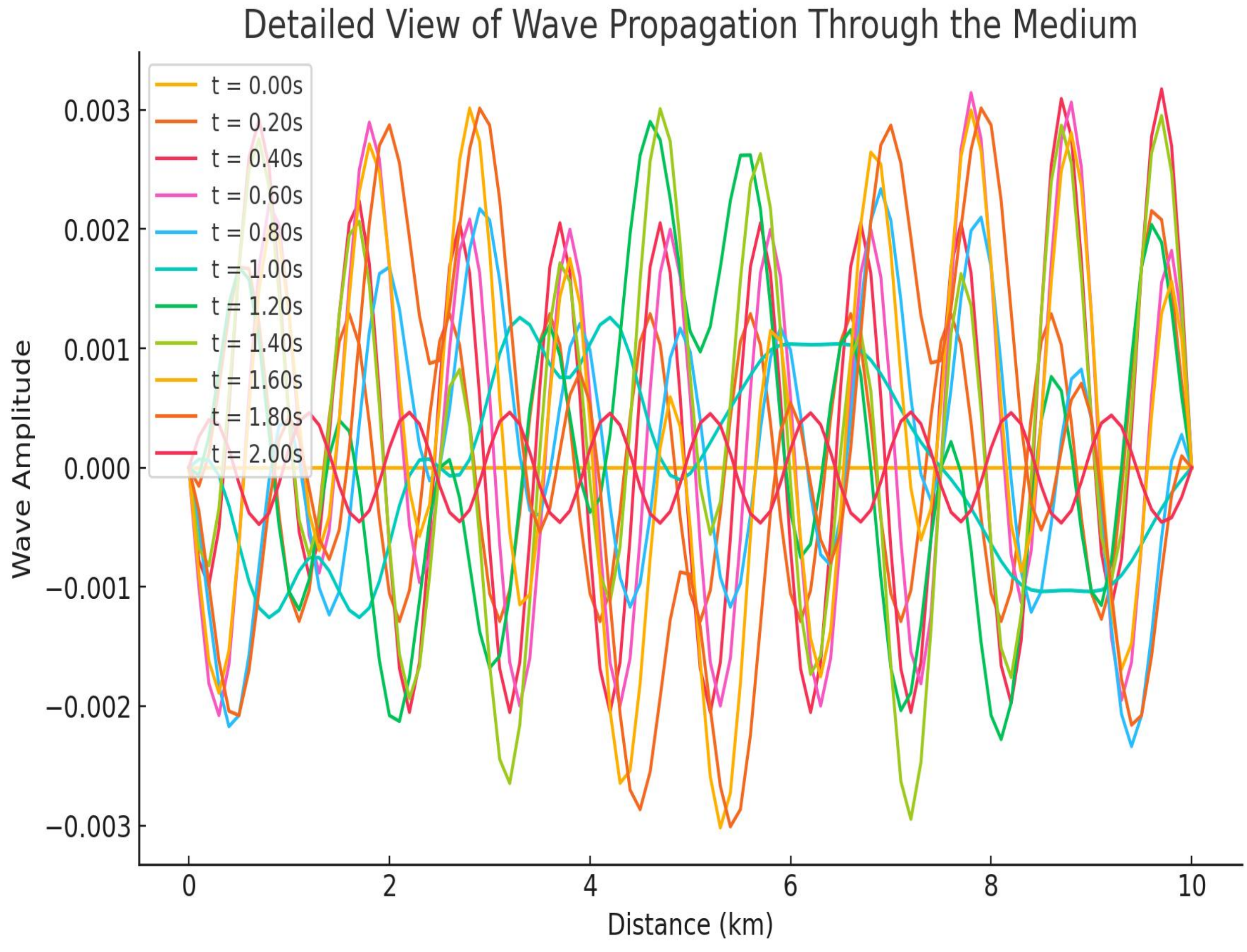

**Figure 2**: Detailed view of wave propagation through the medium.

The phenomena presented in Figure 2 capture the complex behavior of waves as they propagate through different media. As the waves move away from the source, their amplitude decreases, illustrating energy dissipation due to medium resistance and scattering effects this view confirms the research report of Ben-Menahem & Singh (2000); Boore & Thompson (2019). The overlapping waveforms at different times reveal constructive and destructive interference patterns, which are critical in understanding wave behavior in heterogeneous materials.

The variations in wave amplitude and shape over distance indicate interactions with the medium's properties, such as density, elasticity, and potential obstacles. These are consistent with the theory of wave attenuation, where the amplitude diminishes as energy is absorbed by the medium as indicated by Lay & Wallace (1995); Chen et al. (2019); Fichtner et al.(2020). The detailed view provided in Figure 2 helps in identifying the wave's speed and its changing nature over time and space.

## Impact of Varying Amplitude and Frequency on Wave Behavior

Result of the study as indicated in Figure 3 highlights the significant impact of varying amplitude and frequency on wave behavior. Higher amplitudes result in more pronounced peaks and troughs, indicating greater wave energy. Conversely, higher frequencies lead to more rapid oscillations, which could affect the energy dispersion and wave interaction with the medium. The research output as presented in Figure 3 explores the impact of varying amplitude and frequency on wave behavior. Multiple subplots show waveforms generated with different initial amplitudes and frequencies, demonstrating how these parameters affect wave propagation characteristics.

The result of the study presented in Figure 3 is essential for understanding the sensitivity of wave dynamics to initial conditions. Higher amplitudes correspond to waves with more energy, which can travel further before significant attenuation occurs. This is depicted by the larger peaks in the subplots with higher initial amplitudes, showing more pronounced waveforms over the same distance, the outcome of this study confirms the research of Stein & Wysession (2003).

The frequency variation illustrates how different frequencies influence the wave's travel speed and attenuation. Higher frequencies tend to dissipate faster due to increased interactions with the medium's microstructure, leading to quicker energy loss. Conversely, lower frequencies travel further with less attenuation, which is why they are often used in long-range geophysical explorations, the views presented by Shearer (2009) is in line with the output of this study. Result of this study further confirms the research output of Kinsler et al. (2000) including Lighthill (1978); Komatitsch & Tromp (2002); Komatitsch & Tromp (2016); Komatitsch & Tromp (2018); Lay & Wallace (1995) where the scientists state that wave behavior is influenced significantly by initial conditions and medium properties.

**Implication of the Varying Amplitude and Frequency on Wave Behavior**:

The theoretical insights gained in this study have practical implications across various domains, including earthquake hazard assessment, where they can enhance ground motion predictions, and seismic imaging, where they improve subsurface mapping accuracy. Expanding the model to include anisotropic properties will further align theoretical predictions with observed wave behaviors in complex terrains.

Figure 3: Impact of varying amplitude and frequency on wave behavior.

### Practical Implications in Risk Management

The outcome of the numerical analysis presented in this study agrees with the research of Stein & Wysession, (2003) which could enhance groundwork for designing earthquake-resistant infrastructure by providing predictive tools for assessing wave impact on various soil and rock compositions. Future iterations of the model of this research could incorporate empirical data from seismic monitoring stations to improve precision and reliability of disaster risk management.

### Practical Applications of Findings

The practical implications of understanding wave attenuation and amplitude sensitivity to medium properties extended across multiple domains, particularly in earthquake-resistant infrastructure design and risk mitigation strategies.

### Earthquake-Resistant Infrastructure Design

Knowledge of how seismic waves attenuate in different media is critical for designing structures that can withstand earthquakes.

*Material Selection*: Engineers can optimize construction materials to dampen seismic energy effectively, minimizing resonance effects in structures (Stein & Wysession, 2003). For instance, foundations built on high-damping soils can reduce structural oscillations during an earthquake.

- *Localized Building Codes*: The findings support the need for location-specific building codes, accounting for regional soil and geological conditions. Areas with high attenuation coefficients may require stricter design standards to ensure resilience.
- *Seismic Hazard Assessment*: Attenuation modeling aids in estimating the intensity and impact of seismic waves over distances, informing hazard maps and early-warning systems.
- *Urban Planning*: Urban developers can leverage these insights to avoid constructing critical infrastructure in high-attenuation zones, where seismic waves lose energy more slowly, posing greater risks to safety.
- *Energy Infrastructure*: For oil and gas sectors, understanding wave attenuation is pivotal for evaluating the stability of pipelines and storage facilities near fault lines (Carcione et al., 2018; Graves et al., 2016).

### Geophysical and Engineering Applications

- *Subsurface Imaging*: Attenuation data improves the resolution of seismic imaging techniques used in resource exploration and fault mapping (Zhu et al., 2016).
- *Wave-Based Nondestructive Testing*: Findings from this study can refine nondestructive testing methods for detecting cracks and voids in critical structures by analyzing wave attenuation properties.

## Conclusion

1. *Initial Energy and Wave Generation*: The simulation demonstrated a uniform initial wave amplitude of 0.001 m across a 10 km domain (Table 1, Figure 1), confirming a stable source and controlled energy injection. This consistency in initial energy input is critical for predicting seismic wave propagation and designing effective mitigation strategies.

2. *Wave Propagation Dynamics*: Wave propagation analysis revealed significant amplitude variations over distance and time, with values ranging from 0.002 m at 2 km (t=0.1t=0.1 s) to −0.002−0.002 m at 8 km (t=0.25t=0.25 s) (Table 2, Figure 2). These variations underscore the influence of medium properties on wave behavior, which is vital for accurate ground motion prediction in earthquake engineering.

3. *Impact of Amplitude and Frequency*: Parametric analysis showed that wave behavior is highly sensitive to frequency variations. At t=1t=1 s, a frequency increase from 200 Hz to 300 Hz (with constant amplitude 0.2) altered the wave amplitude from 0.1 to 0.2 (Table 3, Figure 3). Tailoring these parameters can optimize wave-based applications in geophysical surveys and environmental monitoring.

4. *Linking Theoretical Models to Practical Applications*: The study's findings provide a foundation for practical applications in earthquake-resistant infrastructure design including material selection and localized building codes (Section 4.5). Furthermore, the insights into wave attenuation and frequency-dependent behavior can enhance subsurface imaging and nondestructive testing methods in geophysical engineering.

## Recommendations

Policy Implications: The understanding of wave dynamics derived from this study can guide the development of improved building codes and disaster preparedness strategies to mitigate seismic event impacts.

## Funding

This research did not receive any specific grant from funding agencies in the public, commercial, or not-for-profit sectors.

## Declaration of Competing Interest

The authors declare that they have no competing interest.

## References

Aki, K., & Richards, P. G. (2002). Quantitative Seismology. Theory and Methods. University Science Books. ISBN: 9780935702965

Allen, R. M., & Melgar, D. (2019). Earthquake early warning: Advances, scientific challenges, and societal needs. *Annual Review of Earth and Planetary Sciences*, 47, 361–388.

Ben-Menahem, A., & Singh, S. J. (2000). Seismic Waves and Sources. Dover Publications.

Boore, D. M., & Thompson, E. M. (2019). Soil amplification and wave resonance effects in seismic design. *Bulletin of the Seismological Society of America*, 109(5), 2351–2371.

Born, M., & Wolf, E. (1999). Principles of Optics. Cambridge University Press.

Carcione, J. M., Picotti, S., & Rubino, J. G. (2018). Wave propagation in partially saturated and fractured media: Theory and numerical simulation. *Geophysics*, 83(1), WA13-WA24.

Chen, W., Shi, Z., & Lu, X. (2019). Seismic wave propagation in anisotropic media: Theoretical and practical insights. *Geophysical Journal International*, 219(3), 1875–1890.

Clayton, R. W., & Engquist, B. (1977). Absorbing Boundary Conditions for Acoustic and Elastic Wave Equations. *Bulletin of the Seismological Society of America*, 67(6), 1529-1540.

Fichtner, A., Lyu, X., & van Driel, M. (2020). Advances in full-waveform inversion for seismic imaging. *Geophysical Journal International*, 222(2), 1135–1150.

Graves, R., Jordan, T. H., Callaghan, S., Deelman, E., Field, E. H., Juve, G., Kim, A., Maechling, P. J., Milner, K. R., Small, P., & Terrel, M. (2016). CyberShake: A physics-based seismic hazard model for southern California. *Seismological Research Letters*, 87(5), 1121–1130.

Kinsler, L. E., Frey, A. R., Coppens, A. B., & Sanders, J. V. (2000). Fundamentals of Acoustics. Wiley.

Komatitsch, D., & Tromp, J. (2002). Spectral-element simulations of global seismic wave propagation. *Geophysical Journal International*, 149(2), 390–412.

Komatitsch, D., & Tromp, J. (2016). Spectral-element methods for large-scale wave simulations. *Geophysical Journal International*, 227(1), 1–30.

Komatitsch, D., & Tromp, J. (2018). Spectral element methods for seismic wave propagation. *Geophysical Journal International*, 213(1), 174–187.

Lay, T., & Wallace, T. C. (1995). Modern Global Seismology. Academic Press. ISBN: 9780127328706

Lee, W. H. K., Kanamori, H., & Jennings, P. C. (2017). Earthquake early warning systems: Lessons from recent applications. *Seismological Research Letters*, 88(4), 1026–1038.

Lighthill, J. (1978). Waves in Fluids. Cambridge University Press.

Nakata, N., Snieder, R., & Behm, M. (2015). Seismic attenuation: Insights from urban environments. *Bulletin of the Seismological Society of America*, 105(5), 2674–2686.

Priolo, E., Seriani, G., & Carcione, J. M. (2020). Attenuation of seismic waves in heterogeneous media. *Physics of the Earth and Planetary Interiors*, 304, 106451.

Qin, Y., Wang, Y., & Zhang, H. (2020). Frequency-dependent attenuation of seismic waves: A review of recent advances. *Earthquake Science*, 33(1), 1–18.

Shearer, P. M. (2009). Introduction to Seismology. Cambridge University Press. ISBN: 9780521882100

Stein, S., & Wysession, M. (2003). An Introduction to Seismology, Earthquakes, and Earth Structure. Wiley-Blackwell. ISBN: 978-0865420786

Stewart, J. P., et al. (2017). Effects of soil amplification on ground motion predictions. *Earthquake Spectra*, 33(1), 23–41.

Takemura, S., Yoshimoto, K., & Furumura, T. (2019). Ground motion prediction using wave propagation models. *Journal of Geophysical Research: Solid Earth*, 124(8), 8716–8732.

Trifunac, M. D. (1971). Surface Motion of a Semi-Cylindrical Alluvial Valley for Incident Plane SH Waves. *Bulletin of the Seismological Society of America*, 61(6), 1633-1658.

Wang, Y., Liu, X., & Zhao, Z. (2021). Modeling wave dispersion in heterogeneous seismic media. *Earthquake Science*, 34(2), 98–115.

Zhu, H., et al. (2020). Seismic imaging for hydrocarbon exploration. *Journal of Geophysical Research: Solid Earth*, 125(8), e2020JB019456.

Zhu, T., Toksöz, M. N., & Nowack, R. L. (2016). Seismic wave scattering and attenuation in heterogeneous media. *Geophysics*, 81(2), WA47-WA56.